\documentclass[]{emulateapj}
\usepackage{longtable}
\usepackage{natbib}
\usepackage{graphicx}
\usepackage{subfigure}
\usepackage{verbatim}
\begin{document}
\title{``Spectroscopic Determination of the Low Redshift Type Ia Supernova Rate from the Sloan Digital Sky Survey"}
\author{K. Simon Krughoff\altaffilmark{1}}
\author{Andrew J. Connolly\altaffilmark{1}}
\author{Joshua Frieman\altaffilmark{2,3}}
\author{Mark SubbaRao\altaffilmark{4,5}}
\author{Gary Kilper\altaffilmark{6}}
\author{Donald P. Schneider\altaffilmark{7}}
\altaffiltext{1}{Department of Astronomy, University of Washington, Box 351580, Seattle, WA 98195-1580, USA}
\altaffiltext{2}{Fermi National Accelerator Laboratory, P.O. Box 500, Batavia, IL 60510, USA}
\altaffiltext{3}{Kavli Institute for Cosmological Physics and Department of Astronomy and Astrophysics, University of Chicago, 5640 South Ellis Avenue, Chicago, IL 60637, USA}
\altaffiltext{4}{Department of Astronomy and Astrophysics, University of Chicago, 5640
South Ellis Avenue, Chicago, IL 60637.}
\altaffiltext{5}{Adler Planetarium and Astronomy Museum, Chicago, IL 60605.}
\altaffiltext{6}{NASA Goddard Space Flight Center, Code 671, Greenbelt, MD, 20771}
\altaffiltext{7}{Department of Astronomy and Astrophysics, 408A Davey Laboratory, University Park, Pennsylvania 16802}
\begin{abstract}
  Supernova rates are directly coupled to high mass stellar birth and evolution.  As such,
  they are one of the few direct measures of the history of cosmic
  stellar evolution. In this paper we describe an probabilistic
  technique for identifying supernovae within spectroscopic samples of galaxies.
  We present a study of 52 type Ia supernovae ranging in age from -14 days
  to +40 days extracted from a parent sample of $\sim$ 350,000 spectra
  from the SDSS DR5.  
  We find a Supernova Rate (SNR) of
  $0.472^{+0.048}_{-0.039}(Systematic)^{+0.081}_{-0.071}(Statistical)SNu$
  at a redshift of $\langle z \rangle = 0.1$.  This value is higher than other values at low redshift at the 
  $1\sigma$, but is consistent at the $3\sigma$ level.  The 52 supernova candidates used in this study comprise
  the third largest sample of supernovae
  used in a type Ia rate determination to date.  In this paper we demonstrate the
  potential for the described approach for detecting supernovae in future spectroscopic surveys.

\end{abstract}
\maketitle
\section{Introduction}

Over the last decade, the study of the properties of type Ia
supernovae (abbreviated SNe in captions and equations) has become one of the fundamental tools used in
observational cosmology. With peak luminosities on the order of M$_V
\sim -$19.3, type Ia supernovae can be observed out to redshifts of
$z > 1$. This enables a broad range of science: 
enrichment of the interstellar medium \citep{sivanandam09}, characterization of
the expansion history of the Universe \citep{schmidt98, perlmutter99},
and understanding the physics of how progenitors transition into supernovae
\citep{mannucci06}.
This scientific potential has
led to the a number of surveys of supernovae in the local and distant
universe \citep{SNF,snls,CSP,miknaitis07,frieman08,CfA}.

The primary method for detecting and characterizing supernovae is through
repeated imaging of the same region of the sky spread over several
months or years. By difference imaging multiple epochs
\citep{alard98}, host galaxies can be subtracted, revealing candidate
supernovae that can then be classified by additional imaging or
spectroscopy. \cite{madgwick03} demonstrated that supernovae can also be
identified in an analogous way from spectroscopic observations by
subtracting the underlying host galaxy spectrum. In this case the
detection and classification of the supernova can be derived from a single
observation; with a direct measure of the redshift, type and age of
the supernova. While less efficient than imaging techniques (as the
observational cost of obtaining the spectroscopic data is much larger)
this approach can be applied serendipitously to any spectroscopic
survey to study, for example, supernovae rates in the local and
distant universe.

\cite{madgwick03} found a local supernovae rate of  $R_{Ia}(z \le
0.25) = 0.4\pm h^2 SNu$\footnote{$SNu = 1/10^{10}L_{B\odot}/100yr$ is the definition for the 
B-band weighted supernova rate used throughout this paper.}, which is consistent with existing measures but was based on only 19
supernovae. In this paper we expand upon this work to undertake a systematic
study of type Ia supernovae as found in the SDSS main galaxy sample. We
consider the statistical and systematic uncertainties that arise when
classifying supernovae from the SDSS spectra, the efficiency of the detection
technique (as a function of redshift, galaxy type and signal-to-noise ratio)
and discuss ways to minimize the impact of systematics due to
misclassification of narrow and broad emission-line
galaxies. Applying these techniques to the 362,431 spectra from the
SDSS DR5 sample we characterize the supernova population and supernova rates for
redshifts $z<0.1$. For calculations dependent on cosmology, we use
$\Omega_m = 0.3$, $\Omega_\Lambda = 0.7$, and $H_o = 70 km s^{-1} Mpc^{-1}$
throughout this paper.

\section{Spectroscopic Samples}

Throughout this paper, we use galaxy samples based on the DR5
\citep{Adelman-McCarthy07} release of the SDSS.
The SDSS spectra are observed through 3'' diameter fibers using a
multi-object spectrograph with a spectral resolution of R $\simeq$
1800 and a wavelength range of 3800 -- 9200
\AA.  Galaxy spectroscopic target selection \citep{strauss02} is based on
imaging from the SDSS camera \citep{gunn98} using the 2.5m telescope
\citep{gunn06} at Apache Point Observatory.  A discussion of the photometric and spectroscopic
pipelines and the spectrophotometric calibration can be found in
\cite{EDR}. The SDSS photometric system is described in detail
in \cite{fukugita96}, \cite{hogg01}, \cite{smith02}, 
\cite{ivezic04}, \cite{tucker06}, and \cite{padmanabhan08}; the filter response curves are given in
\cite{fukugita96}.  The astrometric calibration is described in \cite{pier03}.

For the purpose of this work, we assume that the SDSS spectra have
accurate relative spectrophotometric calibration.  Discussion of the
quality of the SDSS spectrophotometric calibration can be found in
\cite{Adelman-McCarthy07}. For this paper, however, we note that the
rms dispersion in the absolute spectrophotometry for the SDSS is $\sigma =
0.04$ mag based on the differences between photometric fiber magnitudes
and magnitudes synthesized from the SDSS spectroscopy
(\footnote{{\tt http://www.sdss.org/DR6/products/\\spectra/spectrophotometry.html}}) and
that, at worst, the spectra show a 5\% variation in the relative
spectrophotometric calibration at the blue end of the spectrum.  All
spectra are shifted to the restframe and corrected for Galactic
extinction using the dust maps of
\cite{sfd} and applying the reddening spectrum from \cite{ccm} with near ultraviolet updates by \cite{odonnell94}.
The spectra are re-sampled to 3 \AA ~linear bins to match the resolution of
the available supernova templates.

Two galaxy samples were constructed for use in this paper (see Table
\ref{samples} for a synopsis of these samples). We applied the following 
criteria to the 1048960 spectra in the DR5 database:
a redshift limit of $z < 0.2$, a positive
redshift, and the requirement that type derived from the spectrum was
not ``STAR''.  The redshift limit for these data was applied in order
to exclude the high redshift luminous red galaxy (LRG) population.  We
wish to avoid segments of the sample dominated by LRGs as this population
would have significantly different eigen spectra compared to the low redshift
galaxy sample.  

Further we define a statistical sample for use in determining the cosmic supernova rate using the
selection criteria used to create the Pitt-CMU Value Added Catalog (VAC; 
\footnote{{\tt http://nvogre.astro.washington.edu/vac}}).  In the SDSS, each observed spectrum has a 
set of binary flags set either by the data acquisition system or the reduction pipeline.
The statistical sample was assembled using the flags associated with the redshift measurement
of the object.  These redshift flags being set can indicate problems with the spectrum and are for
this reason eliminated from the sample.
See \cite{EDR} for complete descriptions of the flags.  These criteria are:
\begin{enumerate}
\item{Spectroscopically classed as a galaxy (SPECCLASS = 2)}
\item{The following redshift flags are set to zero:\\
       Z\_WARNING\_LOC -- Confidence is low, \\
       Z\_WARNING\_NO\_SPEC -- No spectrum,\\
       Z\_WARNING\_NO\_BLUE -- No blue side spectrum, \\
       and Z\_WARNING\_NO\_RED -- No red side spectrum}
\item{Redshift status flags not set to 0 or 1 (0 = not measured, 1 = failed)}
\item{Redshift confidence level greater than 70\%}
\item{The spectrum is the highest S/N example of the object}
\end{enumerate}
This statistical sample contains 362,431 unique objects. We note that
these criteria are less strict than those employed in building the
SDSS Main Galaxy sample \citep{strauss02} and provides more objects
for analysis. All derived
attributes (e.g.\ apparent magnitudes, absolute
magnitudes, and signal to noise measurements) are taken from the
VAC.
In the redshift range $0.01 < z < 0.2$ the VAC is $4\%$ larger than the Main
sample using the same criteria.  Although the two samples are selected using
different criteria, a contributing factor to the VAC being larger is that the
Main sample is magnitude limited to $m_{fiber_{r}} < 19.0$ where the VAC has no
such limit.

We randomly select a subset of galaxies from the statistical sample to
determine the efficiency and systematics of the supernova detection and
classification method.  No attempt was made to avoid spectra with
supernova contribution already present.  This should not, however,
affect efficiency or systematics estimates, since only $0.025\%$ of
galaxies are expected to contain a supernova.  The model sample is constructed by
randomly selecting two thirds of the runs in the DR5 release which provides
enough galaxies to capture the statistical properties of the statistical sample but
saves compute and storage resources.  This model sample
consists of 234638 galaxies.

For each galaxy in the model sample a type Ia supernova template is added to
the galaxy spectrum (with the supernova age chosen at random from an interval
of -20 to +50 days from peak brightness).  The absolute luminosity of
the supernova is sampled from the distribution of B-band peak brightnesses
given in \cite{dahlen04}.  The luminosity is scaled to the observation
time using the parametrization of the lightcurve introduced in
\cite{goldhaber01}.  We assume a stretch factor $s=1$ for all synthetic supernovae. 
The spectra are then shifted and dimmed based on
the redshift of the host galaxy.  Host galaxy extinction and reddening
is applied by choosing a V-band absorption in magnitudes, A$_V$, from
an exponential distribution, $P(A_V)=e^{(-A_V/m)}$, with the scaling
parameter, $m = 0.33 \pm 0.09$, and using an \cite{ccm} reddening
curve with updates to the near ultraviolet by \cite{odonnell94}.
Galactic reddening is applied using the \cite{sfd} Galactic reddening
maps again assuming a \cite{ccm} with \cite{odonnell94} reddening
curve.  We do not attempt to correlate supernova properties with host
spectral properties.

\section{Method\label{methodsec}}

As described in \cite{madgwick03}, our goal is to identify supernova within
SDSS spectra. In this section we describe our extension of the initial
work of \cite{madgwick03}. We approach this problem by requiring that our
technique must satisfy the following attributes:
\begin{enumerate}
\item Computational  Efficiency: It must be computationally tractable
  to classify spectra in real time.
\item Efficient and Complete Detection: Detection must produce few
  false positives and reject none of the candidates with more than four matched 
  spectral features above the noise in the spectrum.
\item Quantifiable Selection Effects: No method will be exact, but to
  be useful, the method must have uncertainties which can be
  quantified.
\end{enumerate}

Principal component analysis (PCA) of galaxy spectra has shown that,
in the visible regime, spectral energy distributions are inherently
low dimensional \citep{connolly95,folkes96,yip04gal}. The spectra can,
therefore, be represented by a small set of orthogonal components or
eigen spectra. \cite{yip04gal} demonstrated that between eight and
eleven components are sufficient to describe fully the spectral
variation within galaxy spectra in the Sloan Digital Sky Survey
(SDSS;\cite{york00}). This results in a reduction in the complexity or
dimensionality of the spectroscopic data by a factor of 400 and means
that it is computationally tractable to fit host galaxy spectra
jointly with any additional components (e.g.\ template supernovae Ia
spectra). Alternative approaches such as fitting a spectral synthesis
model (e.g. \citep{BC}) can be
computationally expensive due to the breadth of the phase space that
must be searched.

Based on the PCA approach we express each spectrum as a linear
combination of  eigen spectra, $e_{ij}$, scaled by an expansion
coefficient, $a_i$, where $j$ indexes the wavelength bins. In this way the spectrum, $f_j$ is represented by,
\begin{equation}
\hat{f}_j = \sum_{i=0}^{n-1} a_ie_{ij}
\label{eigcomp}
\end{equation}
where $n$ is the number of spectral eigencomponents used. 
If the number of spectral components is less than the
dimensionality of the data, the reconstructed spectrum is said to be a
``lossy'' compression. However, as the PCA is a variance weighted
statistic, this truncated expansion preferentially suppresses the noise
within a spectrum.

\begin{figure}[ht]
\centering
\includegraphics[width=3in]{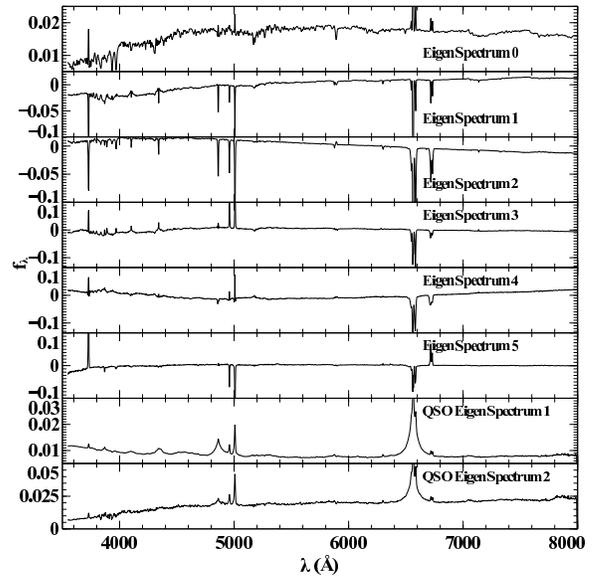}
\caption{First six galaxy eigen spectra and the two QSO eigen spectra used in the fits of our galaxy sample.
  The top most spectrum is the mean spectrum.  The spectra have been
 re-sampled to 3\AA\ linear bins.\label{eig}}
\end{figure}

We use the first eight ($n=8$)
eigen spectra from \cite{yip04gal} to fit the galaxy contribution. 
Figure \ref{eig} shows the first six galaxy eigen spectra and the two
QSO eigen spectra
used in the fitting.
At this level of truncation the
contribution from higher order eigen spectra is typically less than 1\%
of the total flux. 

From this expansion we calculate the error weighted log likelihood ($log(L)$)
value for each spectrum.
\begin{equation}
log(L) = \sum^{m}_{j=0} (f_j - \hat{f}_j)/\sigma_j^2
\label{loglike}
\end{equation}
where $f_j$ and $\hat{f}_j$ are the spectrum flux and reconstructed flux,
respectively, and $\sigma_j$ is the rms error as a function of
wavelength.
\begin{figure}[ht]
\centering
\includegraphics[width=3in]{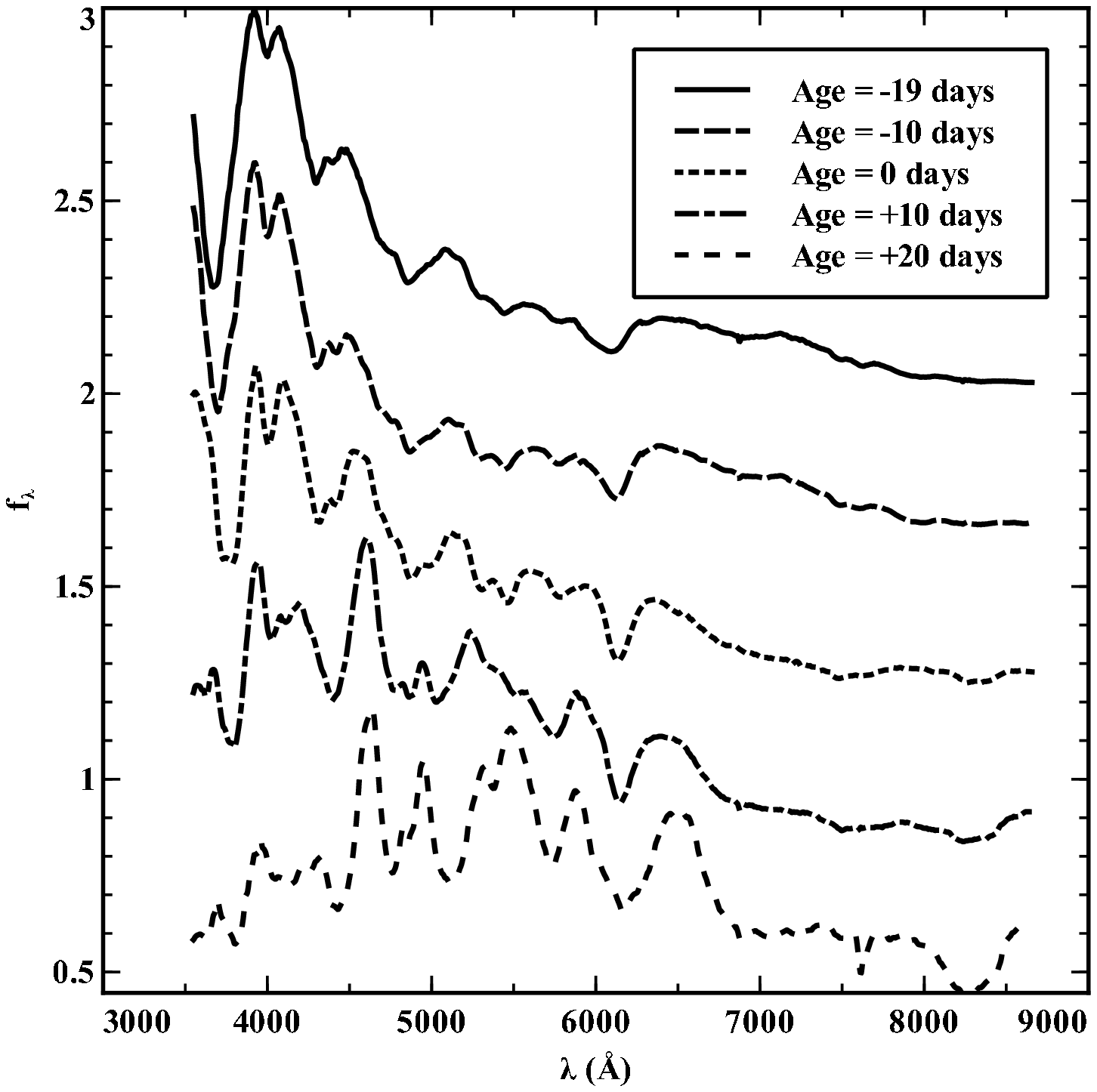}
\caption{Example SN templates from \cite{nugent} at various ages.  Spectra have been offset for clarity.
  \label{sntemps}}
\end{figure}

\begin{figure}[ht]
\centering
\includegraphics[width=3in]{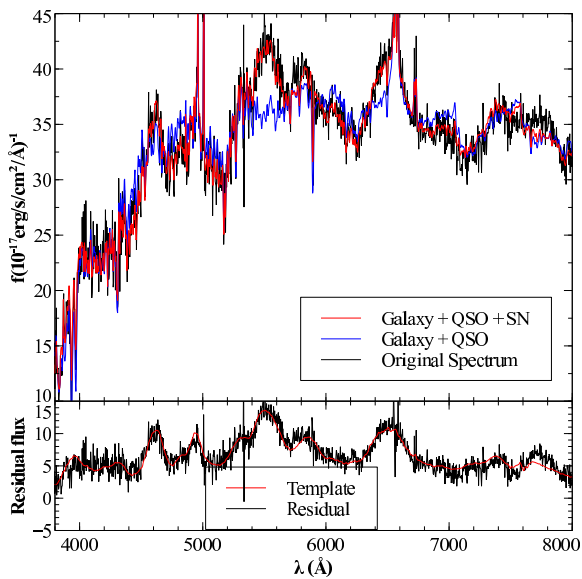}
\caption{ Comparison of the fits with only galaxy eigen spectra (blue
  line) and eigen spectra with a supernova template (red line) for a
  spectrum with significant supernova contribution (SDSS
  J083909.65+072431.6).  The original spectrum is shown in black.
The residual supernova signal from the top pane (Observed-(Galaxy + QSO)).  Note that the flux contribution from the supernova is greater than $50\%$ that of the galaxy.
The residual is shown in black with the best fit supernova template in red.
\label{res}}
\end{figure}

To determine the best fit supernova spectrum,  we refit
each SED using the same eigen spectra together with a series of supernovae templates taken
from \cite{nugent}. These 91 supernova templates range in age, defined as
the rest frame time from peak brightness, from -20
days to +70 days.  Example supernova templates
are shown in Figure \ref{sntemps}.  For each of the supernova templates, we
recalculate the $log(L)$ value and take the minimum $log(L)$ to be the best fit age.  
An example of this procedure is shown in Figure
\ref{res} together with the resulting residuals after subtracting the
galaxy model.

In cases where there is strong nuclear activity, the galaxy
eigen spectra under fit broad emission line features.  Due to a
conspiracy between SiII and OI absorption, type Ia supernova spectra
\citep{branch82} produce an apparent emission feature at $\sim6500$
\AA. This effect creates false positives from confusion between this
absorption feature and broad H$\alpha$ emission in type 1 Seyfert-like
galaxies.  Even though the fraction of AGN is small (5\%-10\%), this
leads to several tens of thousands of false positives.  We therefore
include an additional component in the galaxy model based on the first
two QSO eigen spectra from \cite{yip04qso}. The QSO component fits the
broad and narrow emission which decreases the number of false
positives by an order of magnitude.

\subsection{Selection Criteria for supernova Candidates\label{selectionsec}}

To determine the low redshift supernova rate, we must discriminate
reliably between those galaxy spectra that contain supernovae and
those that do not. To accomplish this we consider two measures of the
significance of a supernova detection: the signal-to-noise of the
residual flux after subtraction of the galaxy and QSO components, and
the goodness of fit of the supernova template measured by Equation \ref{loglike} when the 
flux is estimated using all eigen spectra plus the best fit supernova template.

We define the signal-to-noise  as
\begin{eqnarray}
 \frac{S}{N} = \frac{\sum_{j=4500\mathring{A}}^{6000\mathring{A}} R_{j}}{{\sqrt{\sum_{j=4500\mathring{A}}^{6000\mathring{A}} \sigma^2_j}}},
\end{eqnarray}
where $R_{j} = f_j - \sum^n_{i=0}a_i e_{ij}$ is the residual flux
after subtracting the underlying galaxy and QSO components, and $n$ is
the number of eigen spectra used to model the galaxy and QSO spectra
(in our case the first two QSO components were used). The signal-to-noise is measured over the wavelength
interval 4500\AA\ -- 6000\AA\ in order to exclude contamination of the
flux due to H$\alpha$.

Following the method outlined in \cite{richards04}, we use a
nonparametric Bayes classifier (NBC) to select supernova candidates.
For a general two class system the NBC can be written as:
\begin{equation}
P(C_1|x) = \frac{p(x|C_1)P(C_1)}{p(x|C_1)P(C_1) + p(x|C_2)P(C_2)}
\label{probeq}
\end{equation}
For our problem, $C_1$ is the supernova candidate class ($SN$), and  $C_2$ is the class of galaxies without
supernovae ($GAL$).  The variable $x$ is the two dimensional location
of a source within the $\frac{S}{N}$ -- $log(L)$ space.  The priors $P(C_1)$ and $P(C_2)$ are the probability of 
drawing a galaxy with a supernova ($P(SN)$) and without a supernova ($P(GAL)$) respectively.
We choose $P(SN) = 3/10000$.  This value is consistent with estimates
from historical supernova rate calculations and with the numbers we see when the algorithm is run on the statistical sample.  
This is a two class system so, $P(GAL) = 1 - P(SN)$.

\begin{figure}[ht]
\centering
\subfigure[
The image shows the probability density of the mixture model trained using the sample with synthetic supernovae added. The contours show the 3,4, and 5$\sigma$ confidence regions
for a galaxy containing a supernova given the underlying galaxy distribution using Equation \ref{probeq}.]{
\includegraphics[width=3in]{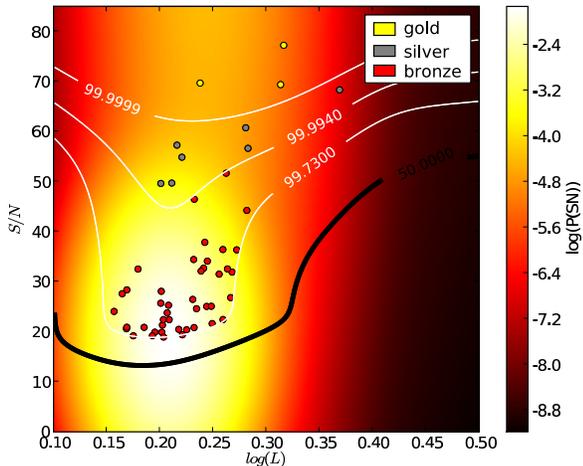}
\label{probaddedplot}}
\subfigure[The image shows the probability density of the mixture model trained using the statistical sample.  The contours are the same as in a.]{
\includegraphics[width=3in]{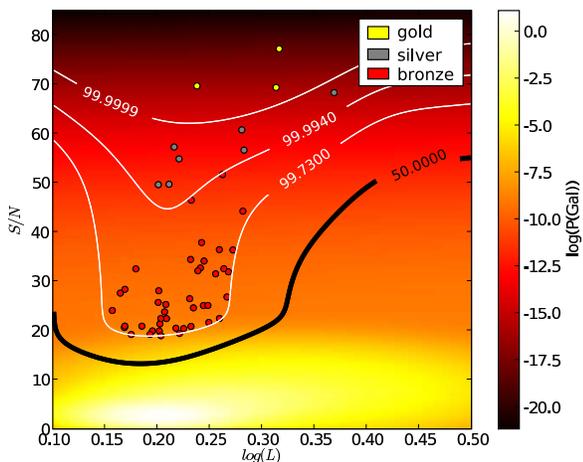}
\label{probgalplot}}
\caption{The training sets for the nonparametric Bayes classifier described in this section. In both panels the color image indicates the probability density of galaxies predicted by the best fit mixture model.  The light colors are regions of high probability density and the darker colors are regions of low probability density.  The probability density is a proxy for the density of points, so regions of high probability density
have more points in the training set.  The points are the supernova candidates that satisfy the criterion that $P(SN|\frac{S}{N}, log(L)) > 0.9973$.  Red points are bronze candidates, gray points are silver candidates, and yellow points are gold candidates.  The thick contour marks the line where a galaxy is equally likely to have a supernova and not have a supernova.\label{probplot}}
\end{figure}

We estimate the relative likelihoods for the NBC using the Statistical
and Model samples described in the previous
section. Figure~\ref{probplot} shows the probability density distributions of the
Statistical and Model samples within the $\frac{S}{N}$ -- $log(L)$
space. White corresponds to high probability density and black
low probability density.  Overlaid on these plots are contours of the value of $P(SN|\frac{S}{N}, log(L))$ corresponding to 3,
4, and 5$\sigma$ confidence levels that a source contains a
supernova.  

Probability density functions for the samples have been
estimated by fitting a mixture of Gaussians using the
FastMix package (\cite{moore99}). The probability $P(SN|x)$ is then
derived from $P(\frac{S}{N}, log(L)|GAL)$, and
$P(\frac{S}{N},log(L)|SN)$.  For our current work we choose to classify a 
galaxy as a supernova candidate if
$P(SN|\frac{S}{N}, log(L)) \ge 0.9973$.  This corresponds to a 3$\sigma$
threshold.  We further define a subset of supernovae the bronze, silver, and
gold candidates (see Section ~\ref{sdss}) using thresholds of 3,
4, and 5 $\sigma$ respectively.

In addition to a cut in $P(SN|\frac{S}{N}, log(L))$, the available supernovae
template models require that all supernovae $-20 < Age < 70$ ( i.e.\ the
template spectra used in the Model Sample span a finite range in
supernova age). To mitigate potential edge effects, we include only
those supernovae with ages $-14 < Age < 40$ within our final candidates.

\begin{figure}[ht]
\centering
\includegraphics[width=3in]{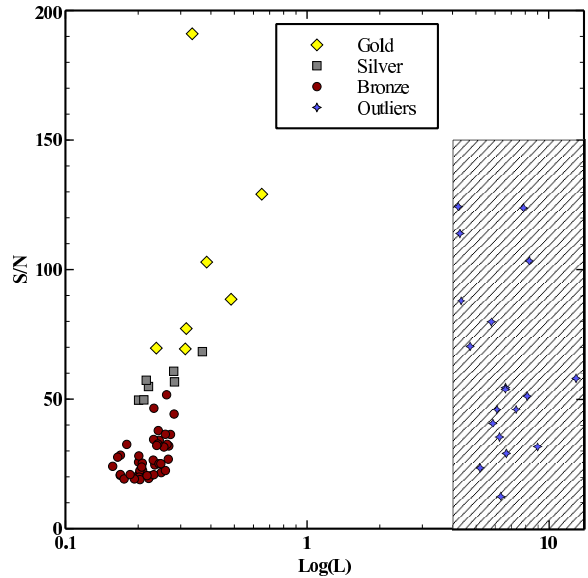}
\caption{The blue circles are sources taken to be supernova
  candidates.  Well separated from this locus is a set of sources
 (red diamonds) that are clear false positives. These arise due to
 some spectra have systematic errors from failures in the
 spectroscopic calibration process, and due to the
 fact that the fitting templates do not completely span the space in
 which extreme galaxy emission line spectra reside. 
 \label{cutplot}}
\end{figure}

Finally, a cut is placed in both $log(L)$ and $\frac{S}{N}$ to avoid
catastrophic failures in the algorithm.  These are generally due to
objects that are not fit well by either galaxy or supernova
templates. By inspection, conservative boundaries are set that exclude
objects with both $log(L) > 4$ and $\frac{S}{N} < 150$ (see
Figure~\ref{cutplot}).  This region is populated by extreme emission
line galaxies where the emission lines influence the
$\chi^2$-statistic such that a good fit is impossible, and objects for
which the spectral calibration failed for one reason or another.
These points are not rejected by the $P(SN|\frac{S}{N}, log(L))$ cut
because neither training set samples this area of parameter space
well.

Together these selection criteria provide a fully automated and
probabilistic approach for selection of supernovae from galaxy spectra.

\section{Simulated Spectra and the Efficiency of Supernova Detection}

The statistical and systematic properties of our supernova detection and
classification technique are determined through a Monte-Carlo
simulation using the model sample described in §2 (and Table 1).

\subsection{Efficiency of Supernova Detection in SDSS Spectra}

Host galaxy properties have a large influence on our algorithm's
ability to detect the resident supernova.  Herein we examine these
properties and discuss their impact on measuring supernova rates.
Efficiencies are calculated by applying the selection criteria
introduced in the previous section, i.e.\, we require that the output
age is $-14 < age < 40$, $P(SN|\frac{S}{N}, log(L)) \ge 0.9973$, and the region with
catastrophic failures is avoided. In the following sections we examine
the impact of host galaxy luminosity, redshift, supernova age, host galaxy
color and signal-to-noise ratio on the efficiency of the method.

\subsection{Efficiency as Functions of Observables}
\begin{figure}[ht]
\centering
\includegraphics[width=3in]{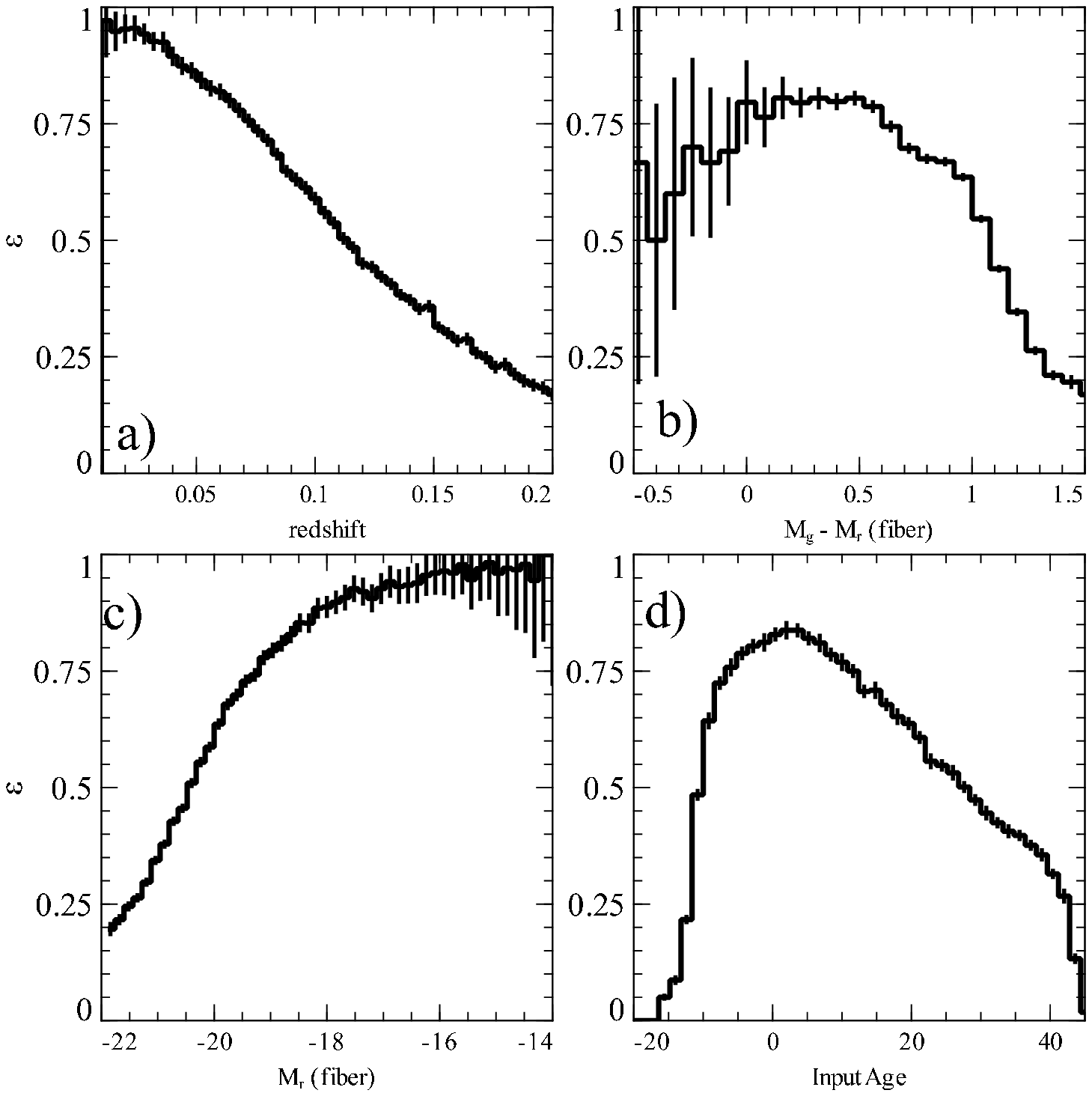}
\caption{Efficiency of the detection algorithm as a function of
  observables:  (a) efficiency as a function of redshift. (b)
  efficiency as a function of galaxy restframe g-r color.  (c) efficiency as a
  function of absolute k-corrected galaxy fiber magnitude in $r$. (d)
  efficiency as a function of input age.\label{multieff}}
\end{figure}

Figure \ref{multieff} shows the efficiency, defined as the ratio of the
number of detected supernovae to the number of input supernovae, as a function of
$r$-band absolute magnitude through the fiber, galaxy restframe color,
age of the supernova and redshift of the host galaxy.

As shown in Figure \ref{multieff}(a), supernova detection efficiency decreases with
redshift (as the signal-to-noise ratio of the underlying spectra will
decrease and the sampled galaxy populations will be intrinsically more
luminous). The efficiency decreases rapidly beyond a redshift of
$z>0.07$, reaching 50\% efficiency at around $z=0.11$. From the
efficiency-signal-to-noise ratio relation shown in Figure \ref{snreff} we
find that this rapid drop in efficiency corresponds to a median per pixel
$r$-band $S/N$ of 50.  This value is calculated by computing the signal-to-noise ratio
for each pixel of the spectrum covered by the r-band filter and taking the median value over 
all pixels.
At $z = 0.1$ the fraction of galaxy spectra with $S/N  > 50$ is
only 3\%.

As expected, the efficiency of detection depends on the intrinsic
luminosity of the host galaxy; as the luminosity of the host galaxy
increases, the efficiency of isolating the supernovae decreases.  The
intrinsic peak luminosity for type Ia supernovae reaches M$_V$ = -19.5
\citep{gallagher08} with under-luminous supernovae scattering to M$_V \simeq
-17.5$. This results in a smooth transition between the regime where
the supernovae dominates the spectrum to where the galaxy host dominates.
Figure \ref{multieff}(c) demonstrates this effect, showing the decrease in the
efficiency of detecting a supernovae remaining at about 75\% to M$_{fiber_r}=18$ and
then dropping quickly to 30\% at M$_{fiber\_r} = -20$ (at which point
the galaxy is contributing about 50\% more light than the peak
brightness supernova).

One might expect the detection of supernovae to be independent of the color
of the galaxy as there are spectral features distributed throughout
the optical spectral range. Figure \ref{multieff}(b) shows, however, a color
dependence in the detection efficiency such that supernova detection in red
galaxies has a lower efficiency than for blue galaxies. This is
due to the correlation between luminosity and intrinsic galaxy
color. If the Model Sample is binned in absolute magnitude and the
efficiencies replotted as a function of restframe color, the efficiency
curve is flat but the normalization of that curve decreases with luminosity.

We also consider the efficiency as a function of supernova age in Figure
\ref{multieff}(d). The integral of this histogram is the control time,
$\tau$, which dictates the temporal window over which a supernova is
detectable.  As shown in Figure \ref{multieff}(d), the impact on efficiency
occurs at early and late times when the supernova luminosity is closest to a
minimum in our templates. For ages $-10< days <+20$, the efficiency of supernova detection
remains above $50\%$.

\subsection{Signal to Noise Ratio}

\begin{figure}[ht]
\centering
\includegraphics[width=3in]{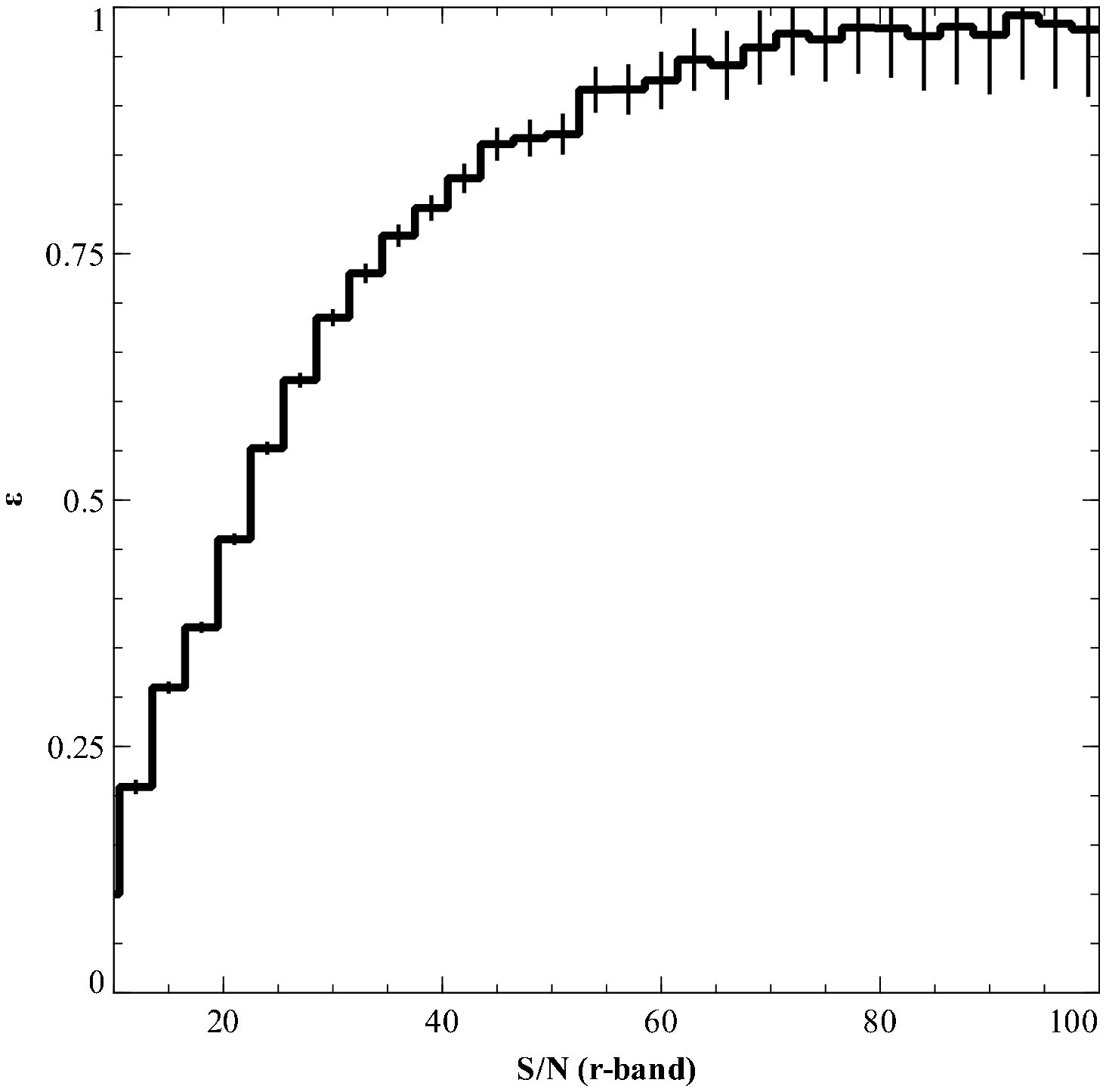}
\caption{Efficiency as a function of signal-to-noise ratio of the measured (galaxy + supernova) spectrum in the $r$-band
  \label{snreff}}
\end{figure}

Based on the results described above, the most important efficiency
indicator for the rate of supernova detection is that of S/N of the measured
spectrum.  The signal-to-noise ratio incorporates all of the dependencies on
galaxy and supernova properties in the efficiency calculation.  In each of
the previous results, the drop in efficiency was related to a decrease
in supernova signal-to-noise ratio (either absolute or relative to the
galaxy luminosity).  We therefore approximate the signal-to-noise ratio of
the measured spectrum (including the supernova) as the median of the signal-to-noise
ratios of the spectral bins in the $r$
passband.  Figure \ref{snreff} shows this efficiency as a function of median
$r$-band S/N.  As expected the efficiency decreases with lower
signal-to-noise ratios, but remains above 50\% down to $S/N = 23$
(which represents 70\% of galaxies in the statistical sample).

\section{Application to SDSS}
\label{sdss}

\begin{figure}[ht]
\centering
\includegraphics[width=3in]{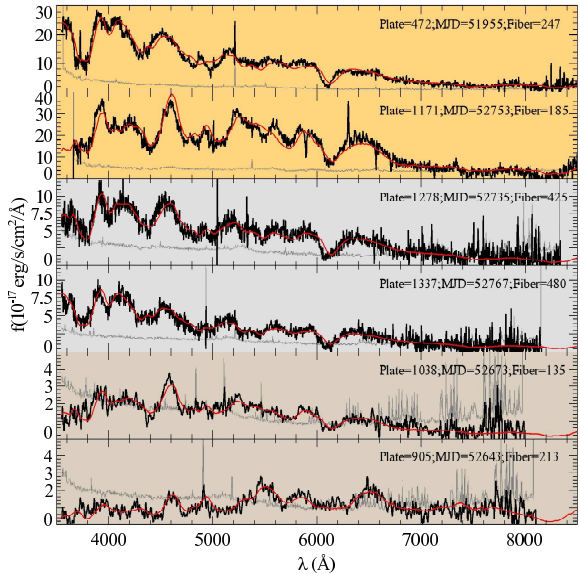}
\caption{
Six examples of supernova candidates selected using the algorithm presented in this paper.  The top two panes are for the "gold" sample, 
the middle two panes are for the "silver" sample, and the bottom two panes are for the "bronze" sample.  The residual after the galaxy component
is subtracted is shown in black.  The best fit template is plotted in red and the rms uncertainty as reported by the pipeline is in gray.
The plate, mjd, and fiber are
provided in the upper right of each pane.  
		  \label{snexamples}}
\end{figure}

The algorithm described in \S 3 and the criteria from \S \ref{selectionsec} were
applied to the SDSS spectra from the DR5 statistical sample.  To
quantify the effectiveness of these techniques we define subsamples of
supernovae as a function of their classification probability. These
supernovae, gold ($P(SN|\frac{S}{N}, log(L))>$0.99999943), silver ($P(SN|\frac{S}{N}, log(L))>$0.999937), and bronze ($P(SN|\frac{S}{N}, log(L))>$0.9973),
are shown in Figure~\ref{cutplot}. Visual inspection of these supernova
candidates shows that: gold candidates have many (greater than four)
spectral features in the residual fit that are coincident with the
best fit template, silver candidates have at least three features that
match the template and exceed the noise in the spectrum, and bronze
class sources have at least three well fit features, but with per pixel signal at the 
level of the per pixel error.  Figure~\ref{snexamples}
shows examples of these three classes. The
bronze candidate spectra were smoothed with a 5 bin tophat filter to show more clearly
how the low frequency signal matches the template spectra.

\begin{figure}[ht]
\centering
\subfigure[Comparison of fiber $u-r$ color distribution for the statistical sample (solid) and supernovae (dotted).  The magnitudes are absolute K-corrected fiber magnitudes.
		  \label{color_hist}]{
\includegraphics[width=3in]{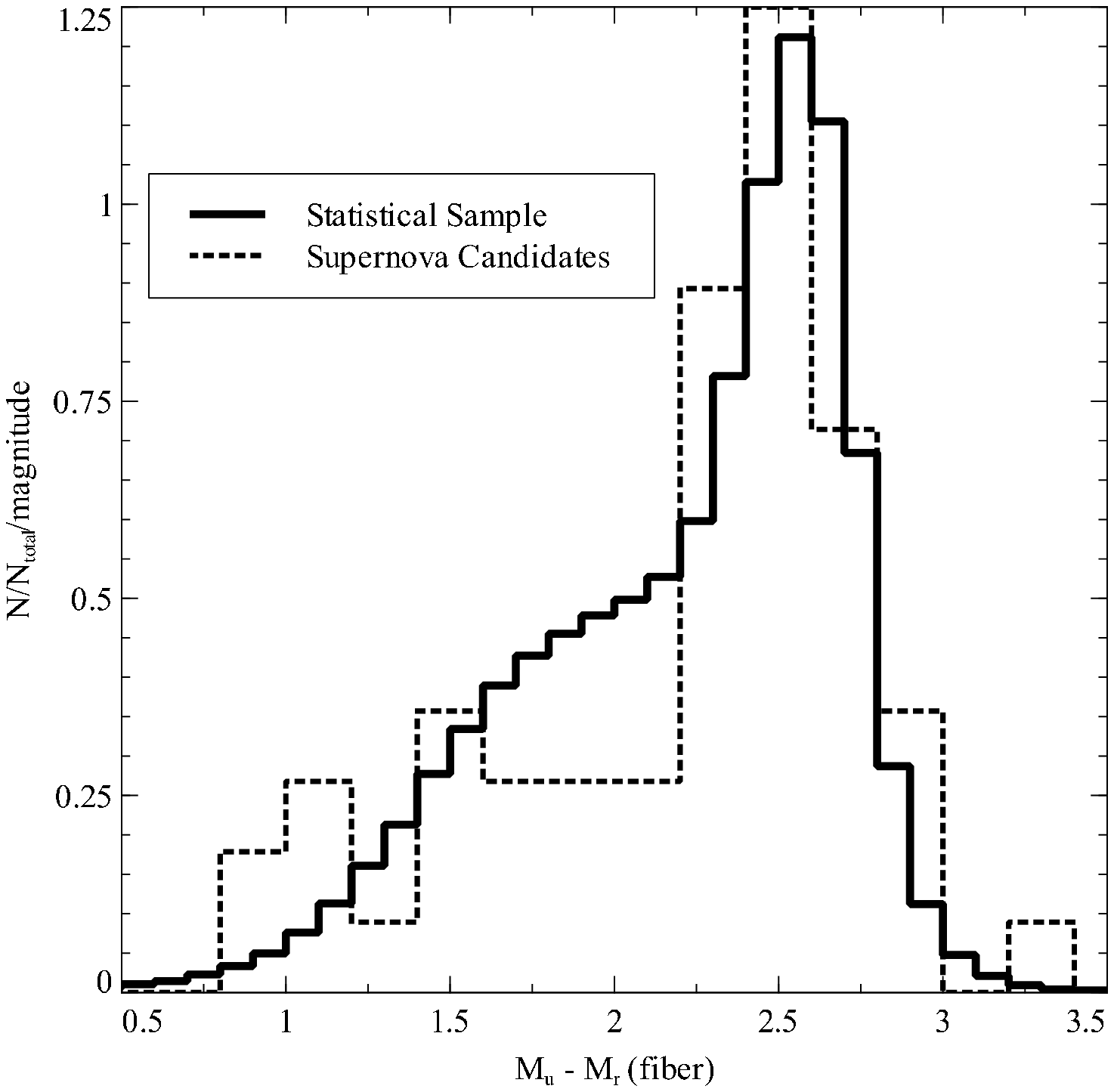}}
\subfigure[Comparison of redshift distribution for the statistical sample (solid) and supernovae (dotted).  
		  \label{z_hist}]{
\includegraphics[width=3in]{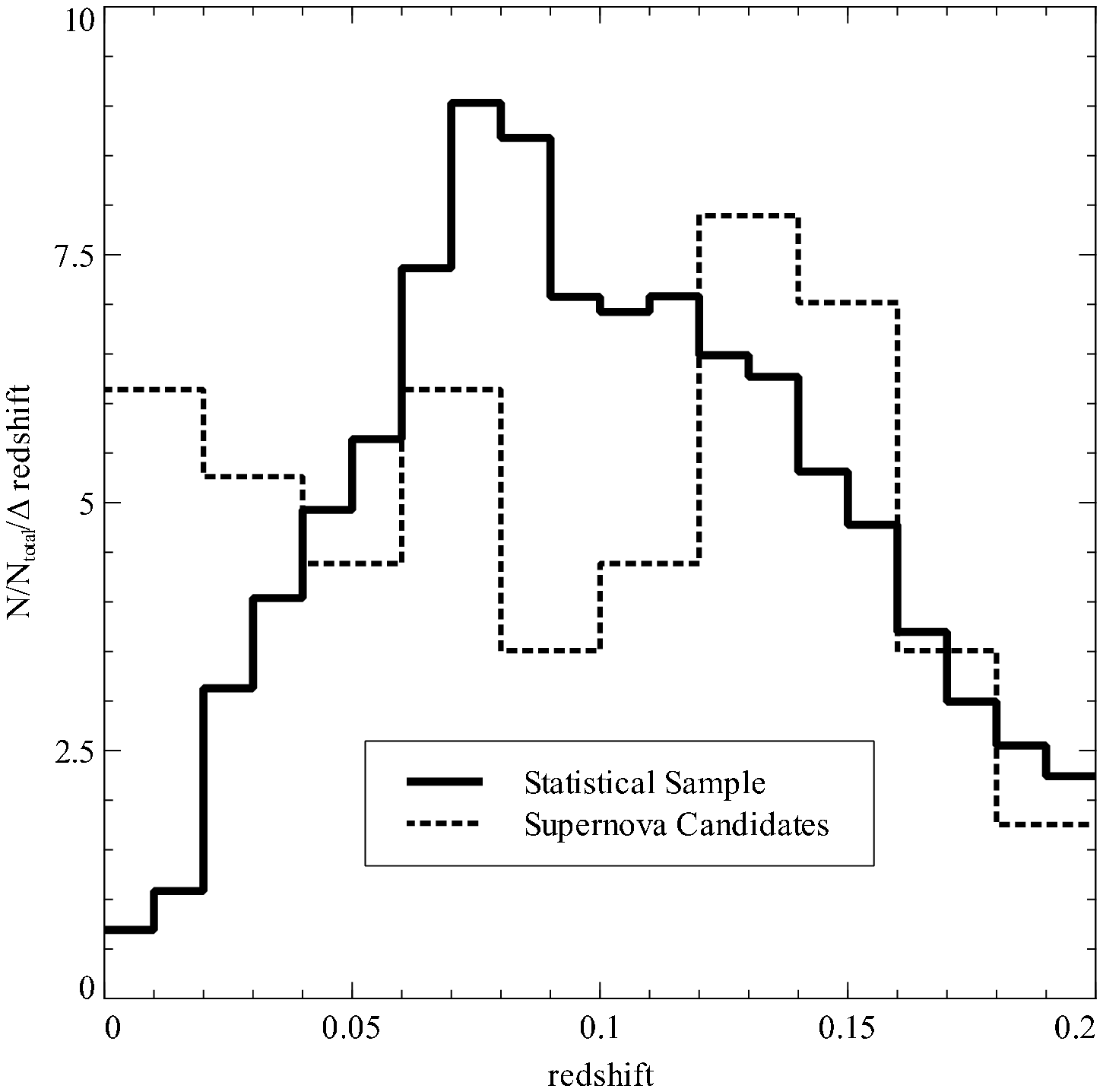}}
\caption{Comparison of supernova candidate distributions to the distribution of the statistical sample.}
\end{figure}

\begin{figure}[ht]
\centering
\includegraphics[width=3in]{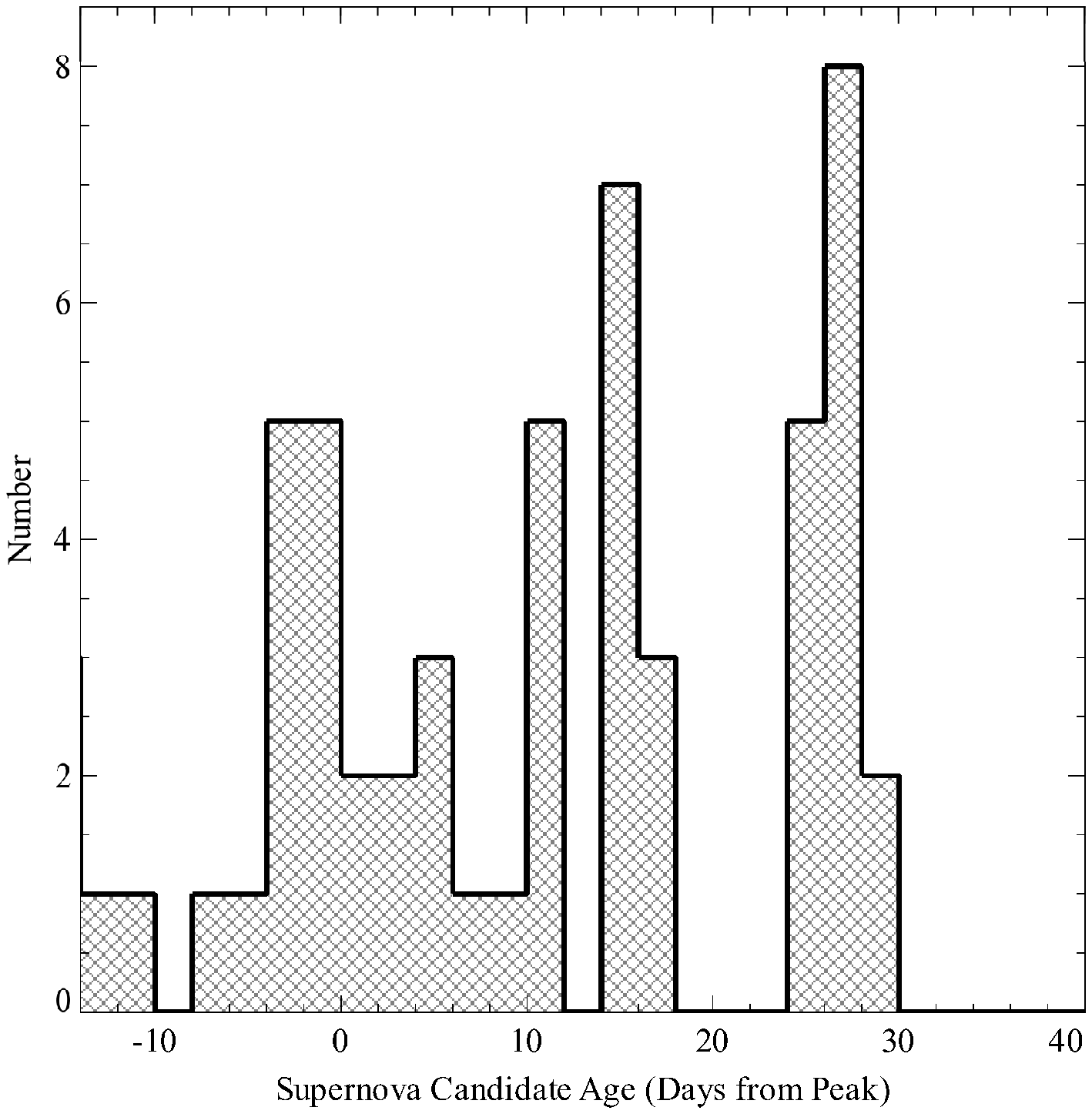}
\caption{Distribution of measured rest frame ages for the 52 SN candidates used in the rate calculation.  
		  \label{age_hist}}
\end{figure}

For reference, we show the distribution in color and redshift for the statistical and candidate samples.
Figure \ref{color_hist} shows that the color distribution is similar for both populations.  The redshift
distributions, however, are quite different. As seen in Figure \ref{z_hist} the candidate supernova population has a much flatter distribution 
than the parent population.  This is likely due simply to the increased efficiency of finding supernovae in nearby galaxies.
Note that the redshift distribution is truncated at $z = 0.2$.
In Figure \ref{age_hist}
we show the distribution of measured ages from peak for the candidate supernovae.  Obviously the number of candidates
is not large enough to fully sample the range of templates, but we do recover supernovae over essentially the entire window of possible ages.


\subsection{Direct Confirmation of supernovae}

\begin{figure}[ht]
\centering
\includegraphics[width=3in]{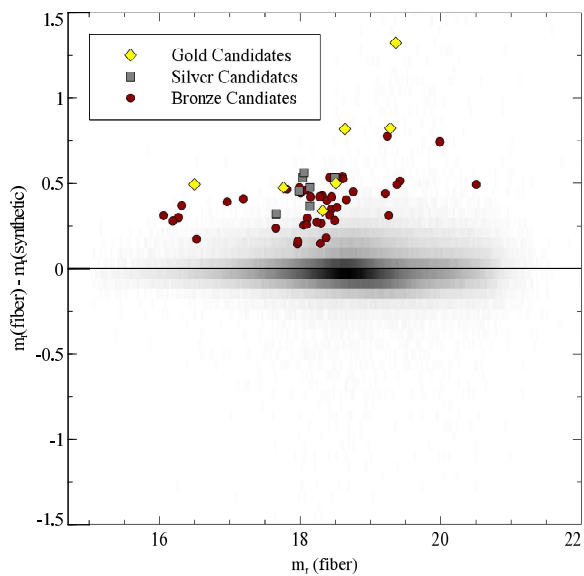}
\caption{
For all detected supernovae, we plot the difference in the $r$-band fiber and synthetic magnitudes against the $r$-band fiber magnitude.  Fiber magnitudes are aperture magnitudes calculated from the photometry.  Synthetic 
magnitudes are calculated by integrating the observed spectrum convolved with fiducial filter curves.  The density plot is the entire SDSS DR5 spectroscopic sample.  Gold candidates are plotted in yellow (diamonds), silver in gray (squares), and bronze in red (circles).  In general, the spectra are brighter than their photometric counterparts, suggesting an
extra flux contribution.  
\label{fibersynth}}
\end{figure}

To validate the accuracy of the detection algorithm we undertake a
number of tests on the derived spectra. The first of these is a
comparison of the fiber magnitudes measured from the photometry with
synthetic magnitudes calculated from the observed spectra.  The fiber
magnitudes are computed by placing a 3" diameter aperture at the
centroid of the seeing convolved object. The synthetic magnitudes are
calculated by integrating the product of the filter transmission
curves and the observed spectrum.  We obtain the fiber and synthetic
magnitudes in the $r$-band for all objects in the full sample. In the
absence of transient events one would expect the two values to be
equivalent.  Figure \ref{fibersynth} shows the difference between the
fiber and synthetic magnitudes as a function of fiber magnitude (for
the statistical sample).  The supernova candidates are plotted as circles.
Yellow circles are gold candidates, gray circles are silver candidates and 
red circles are bronze candidates.

Of the 52 supernova candidates, all show a brightening in the
synthetic magnitude as compared to the fiber magnitude.  This is
consistent with additional flux contribution from transient sources.

\begin{figure}[ht]
\centering
\includegraphics[width=3in]{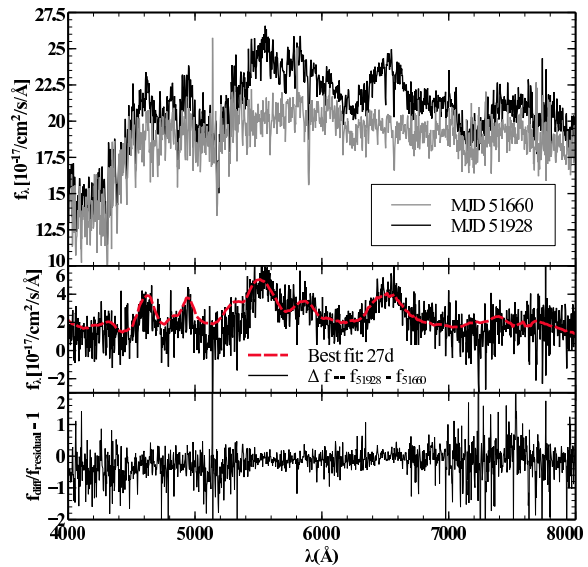}
\caption{
An example of an object with two epochs in spectroscopy of object SDSS j124733.40+000557.1.  In the upper panel Epoch 2 (black) with the SN contribution is obviously brighter than the spectrum taken at the earlier Epoch 1 (gray).  
The middle panel shows the comparison of the difference of Epoch 2 and Epoch 1 (black) to the best fit template as chosen by the SN selection 
algorithm (red dashed).  The best fit template is age 27 days.  The bottom panel is the ratio of the difference spectrum to the galaxy subtracted supernova.  
\label{specdif}}
\end{figure}

Of the 52 supernovae used in the rate calculation, four were observed
spectroscopically on more than one occasion. All four cases show a brightening 
in the supernova containing spectrum.  In all four cases the difference of the two epochs
show similar feature to the residual spectrum from the fitting algorithm.
A typical example, SDSS
J124733.40+000557.1, was observed on MJD 51660 and 268 days later on
MJD 51928.  The supernova was detected in the later epoch with an estimated
age of 27 days after maximum.  This candidate was detected as a member of the bronze 
sample with $P(SN|\frac{S}{N}, log(L)) = 0.9986$.
Given these repeat observations we
subtract the two spectra (which removes the uncertainty in the
modeling of the host galaxy spectrum) and show the two spectra
together in the top pane of Figure \ref{specdif}. In the middle pane we
plot the difference of the flux in the two epochs with the best fit template spectrum scaled
as calculated by the algorithm.  Clearly the difference spectrum and the template agree.
Finally, we plot the ratio of the difference spectrum to the supernova signal from epoch 2.  At
shorter wavelengths it appears that the fitting algorithm is under-subtracting the galaxy contribution.
In this example the under-subtraction is $<5\%$ of the total spectrum flux.  This shows that the fitting algorithm 
does a very good job in this case of superimposed galaxy and supernova signal.

These two pieces of evidence, the brightening of objects in the
spectrum relative to the photometry, and multi-epoch spectroscopic observations
showing supernova signatures indicate strongly that the algorithm is
selecting real supernova signatures.

\section{Type Ia Luminosity Weighted Supernova Rate}

Based on the efficiency calculations in \S 4 and the supernovae identified in
the statistical sample we calculate the supernovae rate for galaxies at a
mean redshift of $z=0.1$.  The luminosity weighted supernova rate
($r_L$) is traditionally defined as:
\begin{equation}
\langle N_{SN}\rangle  = r_L \times S,
\label{snrate}
\end{equation}
where $\langle N_{SN}\rangle$ is the expected number of type Ia supernovae and $S$ is the sum of the probability of detecting a supernova in a
given galaxy, weighted by the luminosity in the B-band, in units of
$L_{\bigodot}$,
\begin{equation}
S = \sum_{i}^{N_{gal}} L_i \epsilon_i \tau_i.
\end{equation}
The sum runs over all galaxies in the statistical sample.  $L_i$ is
the rest frame B-band luminosity in units of $L_{\bigodot}$ and $\tau_i$
is the time period we are sensitive to identify supernova.
The efficiency $\epsilon$ is a function of redshift, apparent
magnitude, luminosity, galaxy type, seeing, and other observables.  As
in \cite{dilday08}, we avoid the complexity involved in modeling the
efficiency as a joint distribution of all characteristics by using the
efficiency as a function of signal-to-noise ratio.  The control
time $\tau$ is then defined as:
\begin{equation}
\tau = \sum_{z = 0.0}^{z = 0.2}\epsilon_z(z)*\int_{t_1}^{t_2} \epsilon_t(t,z) dt
\end{equation}
where $t$ is the age of the supernova with peak brightness
occurring at $t = 0$ and $\epsilon(t)$ is the efficiency as a function
of supernova age.  We use the window $ -14 \le age \le 40$ as this is the interval
over which $\epsilon(t)$ was evaluated.  
We use a redshift averaged value of $\tau$ using the redshift distribution for the 
statistical sample (see Figure \ref{z_hist}) to
obtain the typical detection window for the survey as a whole.  We
evaluate this integral to obtain the fraction of a year over which the
algorithm is sensitive to detecting type Ia supernovae.  For this survey
$\tau = 0.1 $ yr.

We calculate the luminosity weight in two ways.  First, we consider
only the contribution of flux from within the fiber.  This takes a
local view of the luminosity weighting by allowing for only stellar
light encompassed by the fiber.  A second approach is to rescale the
luminosity by the covering factor of the fiber relative to the galaxy
total flux. This covering factor is calculated as:
\begin{equation}
cf = 10^{(M_{model} - M_{fiber})/2.5}.
\end{equation}
In the above equation, the k-corrected, absolute magnitudes may be in
any band but, as the galaxy sample is $r$-band selected, we use the $r$-band 
to calculate the covering factor.  
In some cases, however, the fiber
magnitude is brighter than the model magnitude due to blending of
overlapping galaxies. In order to avoid these issues, we discard
galaxies where $M_{fiber} < M_{model}$.  This situation is encountered in $~0.5\%$
of the galaxy sample.  This results in 360,698 galaxies used in the covering factor 
method of calculating the rate.

In general, using the flux
through the fiber is preferable since there is no dependence on
photometric model fitting.  The model fitting could potentially 
introduce a bias in the covering factor as a function of galaxy 
morphology.  For this reason, we use the fiber-based B-band luminosity
weighted rate for comparison to other published values and in reporting the rate as a function
of galaxy color.

We further subdivide the sample into red and blue subsets using a
color cut of $M_u - M_r = 2.2$ to estimate the supernova rate as a
function of galaxy spectral type.  \cite{strateva01} suggest that
a value of 2.2 gives optimal separation between early and late type galaxies.
We use this value to separate the statistical sample into an early (red) and
late (blue) galaxy sample.  
In this case these values have been calculated using the
measured fiber flux in order to maximize the size of the supernova
sample.

Solving Equation \ref{snrate} for $r_L$, and applying the calculated
values of $S$ and $N_{SN}$, we obtain a value of $r_L$ given a given value
from Equation \ref{probeq}.  In order to get the best estimate of the rate given the variance inherent in 
the sample we select many values of the cutoff: $0.9973 \le P(SN|\frac{S}{N},Log(L)) \le 1.0$.  The lower limit
corresponds to a $3\sigma$ and is the cut used in producing the plots presented in Figures \ref{multieff} and \ref{snreff}.
In practice, new efficiency curves are calculated for each selected value of the probability cut.  We calculate the rate
for 100 different selections of the selection cutoff.  We then find the inverse variance weighted mean of the rates.
This weighting was chosen with the expectation that more stringent cuts would produce less candidates but be less likely to 
include false positives.  If there is no contamination from interlopers, even at the $3\sigma$ level, the inverse variance weighting should have no
impact on the mean. See \S \ref{cuterr} for discussion of systematic error contribution of varying the threshold.  The weighted mean yields $r_L = 0.472 \pm 0.08h^2_{70}SNu$ 
which assumes that the supernova rate is proportional
to the rest frame B-band galaxy luminosity.  The statistical error is the central
$68.3\%$ Poisson confidence interval assuming the median number of 
supernova candidates (45) for the 100 data points. 
The supernova rate values and details of the errors are
summarized in Table \ref{snval}. 
Note that the total number of supernovae listed in the table in column 6 differs
from that listed in Table \ref{samples}.  This is due solely to the fact that the $-14 < Age(Days) < 40 $  age cut was not applied to 
the candidates selected from the Statistical Sample.
The supernova rates for the red and blue
samples are $r_L(red) = 0.379 \pm 0.08$ and $r_L(blue) = 0.394 \pm 0.12$, respectively.
As seen in other samples, the luminosity weighted rate from blue galaxies is higher
than that from red galaxies, however they are consistent within errors.

\subsection{Systematic Errors in Supernovae Rates}

We consider potential systematics in our derived values of $r_L$.

\subsubsection{Color corrections to a B-band luminosity} 
We compare the derived
B luminosity for the host galaxies using two relations for color
transformations between the SDSS and Johnson B passbands.  These color
corrections are taken from Lupton (2005)\footnote{\tt
http://www.sdss.org/dr7/algorithms/ sdssUBVRITransform.htmlk\#Lupton2005}
and \cite{jester05}. Specifically the corrections are:
\begin{equation}
B      =    g + 0.33*(g-r) + 0.20 Jester et al. (2005)
\end{equation}
\begin{equation}
B = g + 0.3130*(g - r) + 0.2271
\end{equation}

The difference between these solutions amounts to
$\pm3\%$ in the derived luminosities.

\subsubsection{ Type Ib/c Contamination} 
\begin{figure}[ht]
\centering
\includegraphics[width=3in]{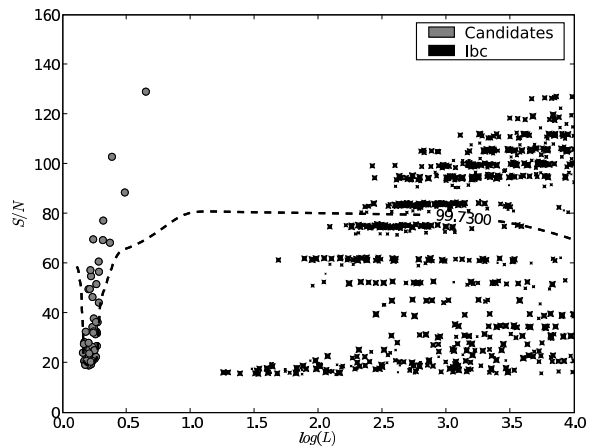}
\caption{
Type Ia candidates (circles) plotted with the results of fitting type Ib/c spectra with type Ia templates (stars).  The size of the point
indicates the age of the type Ib/c candidate.  The dotted line is the $3 \sigma$ threshold applied when selecting candidates.  Many of the candidates are ruled out by the threshold.
\label{bccontam}}
\end{figure}

Since we use only type Ia templates for candidate selection, it is possible that 
type Ib/c supernovae could be a source of false positives.  To test how sensitive this 
algorithm is to type Ib/c interlopers, we simulate perfect galaxy subtraction by
fitting our set of type Ia templates directly to type Ib/c templates (\cite{levan05}) at various 
signal to noise levels.  We can then plot the results of the fit directly on in the same space as
that used for candidate selection.  Figure \ref{bccontam} shows where the type Ib/c spectra fall on 
the $Log(L) -- \frac{S}{N}$ space.  Many are excluded by the $3\sigma$ confidence threshold (dashed line).  
Those that do pass are well separated from the candidates in our sample.  If we had candidates in the
region of the discriminate space occupied by the type Ib/c spectra, we would need to account for the possibility
of contamination.  The locus of our sample is very well separated from that of the type Ib/c candidates, our sample shows
no evidence of contamination by type Ib/c supernovae.

\subsubsection{Efficiency Uncertainty} 
The efficiency histogram has an intrinsic uncertainty due
to
shot noise (based on the number of galaxies in each bin). 
Propagating Poisson errors on the signal-to-noise ratio efficiency shows an uncertainty of not more than 2\% in the rate calculation when $1\sigma$ errors are considered.  We adopt a value of 2\% for the contribution of uncertainty of efficiency.

\subsubsection{Uncertainty in Threshold Cut \label{cuterr}}
The value of Equation \ref{probeq} gives a good discriminant for distinguishing supernova candidates, but the choice of 
the cutoff value influences the calculated value of the rate.  To get a handle on how much this
affects the rate value we calculate rates using 100 different choices for the cutoff.
We then calculate the inverse variance weighted mean of the rate give the 100 data points.  We then 
quantify the scatter by calculating the standard deviation of the sample greater than the mean and the 
standard deviation of the sample less than the mean.  This analysis gives a positive scatter of 9.5\% and a 
negative scatter of 7.5\%.  The fact that the scatter to higher values of the rate is larger than to lower values
suggests that there is some contamination from interlopers, but that when several values of the cutoff are taken, the
effect on the measured rate is small.

Adding these contributions in quadrature this estimate of
the systematic error corresponds to $^{+10.2\%}_{-8.3\%}$.

\section{Comparison to Models}

Many type Ia supernova rate (SNR$_{Ia}$) measures have recently appeared in the
literature, often at higher redshift than previously sampled.  Measures at
high redshift ($z>1$) are generally given in units of the co-moving
volumetric element.  In order to directly compare the rate derived in
this paper with those in the literature, we convert luminosity weighted rates
to volumetric rates using the B-band luminosity density.
As in \cite{horesh08} we use the redshift dependent form $j_B(z) =
(1.03 + 1.76 \times z) \times 10^8 L_\odot Mpc^{-3}$ from \cite{botticella08}.
Table \ref{snrates} is a collation
of SNR$_{Ia}$ from the literature.  We have ordered them 
in mean redshift to facilitate comparison of values at similar distances.

The cosmic supernova rate is notoriously difficult to measure as the number of
events is usually small, the systematics are numerous and hard to correct for, and
even the data on a per object basis can be less than ideal because of their 
transient nature.  All these factors contribute to large errors, which
make differentiation between different predictions difficult.  

Two formulations of the analytic SNR$_{Ia}$ are typically used.  The delay time distribution (DTD) formulation
can be expressed using
the the notation of \cite{greggio05}, the SNR$_{Ia}$(t) is defined:
\begin{equation}
SNR_{Ia}(t) = k_\alpha \int_{\tau_i}^{min(t,\tau_x)} A_{Ia}(t - \tau) \dot{\rho}(t - \tau) f_{Ia}(\tau) d\tau
\end{equation}
where $k_\alpha$ is the normalization of the initial mass function (taken to be 2.83 for the Salpeter IMF), 
$A_{Ia}$ is the efficiency of the progenitor channel, $\dot{\rho}$ is the cosmic star formation rate, and $f_{Ia}$ is
the distribution of times between birth and explosion also denoted DTD$_{Ia}$.  $A_{Ia}$ is typically taken to be constant with time, but 
in general can evolve with the stellar population.

The ``A+B" model formalized in \cite{sullivan06} scales a tardy component and a prompt component based on the integrated stellar
mass build-up and instantaneous
star formation rate, respectively.  Using the notation of \cite{hopkins06},
\begin{equation}
SNR_{Ia} = A\rho_*(t) + B\dot{\rho_*}(t).
\end{equation}
Both the DTD and ``A+B" methods can have short and long timescale contributions to the total
SNR$_{Ia}$ at any given time, but the DTD formulation allows much more latitude in the spectrum of the delay times.  \cite{hopkins06}
modified the ``A+B" model by allowing for a delta function delay time in the prompt component effectively setting a characteristic
delay time.  They found that they could set the tardy component to zero, yielding:
\begin{equation}
SNR_{Ia} = B\dot{\rho}(t - \tau)
\end{equation}
where $\tau$ is the delay time and is $~3Gyr$.

Most DTD models rely on white dwarf binary systems as
progenitors of type Ia supernovae.  However, it is not yet clear what the
companion star is in these binary systems
\citep{branch95,parthasarathy07}.  Many groups have attempted to
constrain DTDs \citep[see][and others]{valiante09,ruiter09,totani08,hachisu08,greggio08}.  
Recent measurements and models of the type Ia DTD imply a
featureless power law.  The two main channels for type Ia supernova
progenitors are the single degenerate (SD; white dwarf with
non-degenerate star \citep{whelan73}) and double degenerate (DD; white
dwarf binary\citep{iben84}) scenarios.  Models of the DD path show
that a featureless power law is in agreement with the predicted DTD
from the DD contribution (\cite{totani08} and references therein).
Recently, it was shown that the SD path can produce a power law DTD
when both white dwarf + main sequence and white dwarf + red giant
systems are considered \citep{hachisu08}.  Direct measurements also
support power law DTDs for type Ia supernovae \citep{totani08,pritchet08}.
The power law DTD is attractive because of it's simplicity and the
fact that it has contribution from both prompt and tardy components.

There is much discussion about which model for the evolution of the cosmic type Ia
supernova rate is the most appropriate, but even for power law DTD models exponent of the power law $\alpha$ is under debate.  Theoretical
values seem to point to a steep power law $\alpha \simeq -1.0$ from both SD
and DD channels \citep{hachisu08,yungelson00,greggio05}.  The value
as measured from data in \citep{totani08} agrees well with $\alpha = -1.0$, although
\cite{pritchet08} find a shallower power law with $\alpha \simeq -0.5$.

\begin{figure}[ht]
\centering
\includegraphics[angle=270, width=3in]{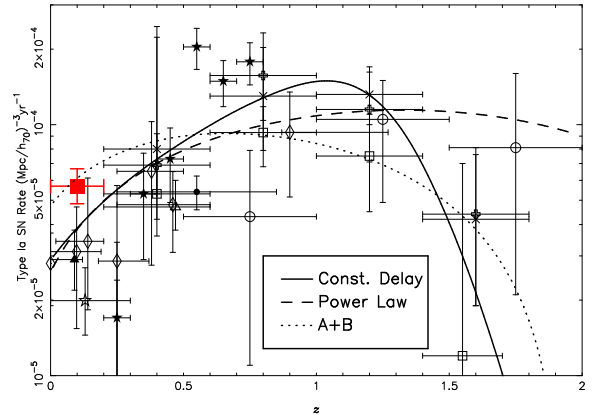}
\caption{ Latest volumetric supernova rates with three models:
a constant delay time of 3Gyr, an A+B model and a 
power law model with $\alpha = -1.0$.  The value from this work with
error bars is plotted in red.
   \label{snr}}
\end{figure}

Changing from cosmic time to redshift and substituting a power law for the DTD model, we produce the following formula:
\begin{equation}
SNR_{Ia}(z)  = k_a\int_{0.3}^{t(z)} A_{Ia}\dot{\rho}_*(t(z) - \tau)\tau^{\alpha}d\tau
\label{lpsnr}
\end{equation}
where $t(z)$ is the cosmic time at redshift $z$, $\dot{\rho}_*$ is
the cosmic star formation rate, $t$ is the delay time and $\alpha$
is the slope of the power law.
The product of $k_aA_{Ia}$
is the constant for which we fit. 
It should be noted that we are not attempting to pin down any physical constants to higher precision
than is already reported in the literature.  The error bars used are statistical in all cases.  No attempt
has been made to cull the data of unreliable data points as the model fitting is intended to be illustrative
rather than diagnostic.
Figure \ref{snr} shows the data from the literature fit using the above models with 
$\alpha = -1$ (dashed line).  This model, supported by theory and data, over predicts the rate
at both low and high redshift, except for the point at $z = 1.2$ from \cite{poznanski07}, who question
the sharp decline in SNR$_{Ia}$ after $z = 1.5$.

The A+B model appears to be too flat to account for both the rise in rate from $z = 0$ to
$z \simeq 1.0$ and the decline after $z = 1.5$.  Since a large part of the best fit A+B model
is contributed by the prompt ($\dot\rho_*$) component which does not peak until $z \simeq 2$, it
is not surprising that it continues to rise at high redshift.
As mentioned above, \cite{hopkins06} noticed that a modified A+B ($SNR =
A\rho_*(t) + B\dot\rho_*(\tau - \tau_{Ia})$; \cite{scannapieco05})
type model with $A=0$ and a delta function delay time distribution do
just as well in describing the observed downturn at $z \simeq 1$.
Figure \ref{snr} shows the best fit \cite{scannapieco05} model with
fixed delay time $\tau_{Ia} = 3Gyr$.  In this scenario, the accrued
stellar mass contributes nothing to the cosmic SNR.  Instead, type Ia
supernovae arise after a fixed waiting period after a star formation event.  This
model is certainly an over simplification as a broad distribution of
assembly times surely exist for both the SD and DD scenarios.

The result from this work is higher than other measurements at low redshift, but is 
within the error bars of all of them.  A possible explanation of the higher measurement
is that systematic effects tend to drive efficiency down.  This leads to lower predicted
rate measurements if these effects are not taken into account.  This argues for a greater level
of correction in the measurement from this paper as compared to others for $z<0.3$.  Despite
being higher than other measurements, our value is still lower than two of the three models
explored in this paper.  The model that is closest to the low redshift values is that of
the constant delta function delay time model.

Current models for the DTD of type Ia supernovae show good agreement with the
data despite DTDs as different as power law and delta functions. The
ability for observed supernova rates to distinguish between these models will
depend on reliable SNR measurements at $z = 2.0$.

\section{Conclusions}

From a sample of 362,431 galaxies we detect 57 (52 of which meet all criterion
for calculating the supernova rate) type Ia supernovae within
SDSS spectroscopic observations by modeling and subtracting the host
galaxy component. Extensive Monte Carlo simulations of the
efficiencies and systematics present in these samples are used to determine a
type Ia supernova rate of 0.472 $h^2_{70} SNu$ at a mean redshift of
$z=0.1$. 
The measurement published here is higher than others at 
$z<0.2$, but agrees within 1 sigma statistical errors with the other published values. It is interesting to note that
the bias for this study is strongly toward finding supernovae in the central parts of galaxies. This is opposite
to the bias of difference imaging type surveys which are slightly biased away from finding supernovae in the central parts of galaxies
due to the brightness of the cores as well as dust extinction.

Despite
our measurement being slightly higher than others at similar redshift, there is still a significant rise
in the type Ia rate from our value at $z = 0.1$ to redshift of unity. 
One possible explanation for this trend is that we were able to incorporate many 
contributors to the efficiency calculation including realistic estimates of host 
extinction and peak supernova luminosity.  Almost any modeled physical effect will tend
to drive efficiency down resulting in higher recovered rate measurements. 
This study is based on one of the largest samples of supernovae in the local
universe. Subdividing the galaxy population into red and blue
components, we find no evidence for a difference in supernova rate as a
function of host galaxy rest frame color within errors.

The success of this study demonstrates that spectroscopic surveys can
be used to identify and classify supernova and supernova rates in the local and
distant universe in a serendipitous manner.  
The next generation of 
wide field spectroscopic surveys such as BOSS \citep{boss} we will
have the potential to detect thousands of supernova in a much larger volume than that sampled by the SDSS
spectroscopic sample.

\section{Acknowledgments}

Funding for the SDSS and SDSS-II has been provided by the Alfred P. Sloan Foundation, the Participating Institutions, the National Science Foundation, the U.S. Department of Energy, the National Aeronautics and Space Administration, the Japanese Monbukagakusho, the Max Planck Society, and the Higher Education Funding Council for England. The SDSS Web Site is http://www.sdss.org/.

The SDSS is managed by the Astrophysical Research Consortium for the Participating Institutions. The Participating Institutions are the American Museum of Natural History, Astrophysical Institute Potsdam, University of Basel, University of Cambridge, Case Western Reserve University, University of Chicago, Drexel University, Fermilab, the Institute for Advanced Study, the Japan Participation Group, Johns Hopkins University, the Joint Institute for Nuclear Astrophysics, the Kavli Institute for Particle Astrophysics and Cosmology, the Korean Scientist Group, the Chinese Academy of Sciences (LAMOST), Los Alamos National Laboratory, the Max-Planck-Institute for Astronomy (MPIA), the Max-Planck-Institute for Astrophysics (MPA), New Mexico State University, Ohio State University, University of Pittsburgh, University of Portsmouth, Princeton University, the United States Naval Observatory, and the University of Washington.
\clearpage
\bibliographystyle{apj}
\bibliography{refs}
\clearpage
\begin{table}
\begin{center}
\begin{tabular}[t]{|l|c|c|c|}
\hline
Sample & Number of Spectra & N$_{SNe}$ & Notes\\
\hline
Statistical Sample & 362,431 & 52 & G=7,S=7,B=43\\
Model Sample & 234638 & N/A & \\
\hline
\end{tabular}
\caption{The samples used in this paper.  The first column gives the name of the sample as it is used in the
text.  Column 2 gives the sample size.  In Column 3 we report the number of supernova candidates in the sample.  This is the total
number of candidates.  The number listed here includes all selection criteria listed in \S \ref{selectionsec}. This number increases to 57 if no age cut is applied.  Any notes are
given in Column 4.  In particular, the number of gold (G), silver (S), and bronze (B) candidates are noted. For a description of how the 
gold, silver and bronze candidates are defined, see \S \ref{selectionsec}.
\label{samples}
}
\end{center}
\end{table}
\clearpage

\begin{table}
\begin{center}
\begin{tabular}[t]{|l|c|c|c|c|c|}
\hline
Method & Value (SNu) & Sys. Error (SNu)& Stat. Error (SNu) & $\overline{N_{SN}}$ & $N_{SN}$\\
\hline
Covering Factor&0.454&$^{+0.046}_{-0.038}$&$^{+0.079}_{-0.068}$&44&51\\
Fiber&0.472&$^{+0.048}_{-0.039}$&$^{+0.081}_{-0.070}$&45&52\\
Red&0.379&$^{+0.039}_{-0.031}$&$^{+0.080}_{-0.067}$&32&36\\
Blue&0.394&$^{+0.040}_{-0.033}$&$^{+0.142}_{-0.108}$&13&16\\
\hline
\end{tabular}
\caption{The results from applying the algorithm to the SDSS Statistical Sample. Column 1 describes the method used, the value of the 
rate is reported in Column 2,  systematic and statistical errors are in Columns 3 and 4 respectively.  Column 5 contains the median
number of supernova candidates in the samples used to calculate the weighted mean of the rate. Finally, we report the number
of candidates used in each of the methods in Column 6.
\label{snval}
}
\end{center}
\end{table}
\clearpage
\begin{center}
\begin{longtable}[]{|c|c|c|c|c|c|c|l|}
\caption[Supernova Rates]{Compilation of volumetric supernova rates from the literature.  Column 1 gives the minimum redshift, column 2 is the maximum redshift, column 3 is the average redshift, column 4 is the volumetric rate, column 5 is the systematic error, column 6 is the statistical error, column 7 is the number of SNe used in the measurement and column 8 contains the reference code (see footnote 4).}\label{snrates}\\ 

\hline 
   \multicolumn{1}{|c|}{\textbf{Min(z)}} &
   \multicolumn{1}{|c|}{\textbf{Max(z)}} &
   \multicolumn{1}{|c|}{$\mathbf{<z>}$} &
   \multicolumn{1}{|c|}{\textbf{Rate}\footnotemark[3]} &
   \multicolumn{1}{|c|}{\textbf{Stat.}} & 
   \multicolumn{1}{|c|}{\textbf{Syst.}} & 
   \multicolumn{1}{|c|}{$\mathbf{N_{SN}}$} &
   \multicolumn{1}{|c|}{\textbf{Reference}\footnotemark[4]} \\ \hline \hline
\endfirsthead

\hline
   \multicolumn{1}{|c|}{\textbf{Min(z)}} &
   \multicolumn{1}{|c|}{\textbf{Max(z)}} &
   \multicolumn{1}{|c|}{$\mathbf{<z>}$} &
   \multicolumn{1}{|c|}{\textbf{Rate}\footnotemark[3]} &
   \multicolumn{1}{|c|}{\textbf{Stat.}} &
   \multicolumn{1}{|c|}{\textbf{Syst.}} &
   \multicolumn{1}{|c|}{$\mathbf{N_{SN}}$} &
   \multicolumn{1}{|c|}{\textbf{Reference}\footnotemark[4]} \\ 
\endhead

   \multicolumn{8}{l}{\textbf{Continued on next page \ldots}}\\
\endfoot

\hline \hline
\endlastfoot

0.0 & 0.0 & 0.0 & 2.8 & $^{+0.9}_{-0.9}$ &   & 70 & g,e \\
\hline
0.0 & 0.12 & 0.09 & 2.9 & $^{+0.9}_{-0.7}$ & $^{+0.2}_{-0.0}$ & 17 & f \\
\hline
0.0 & 0.19 & 0.098 & 3.12 & $^{+1.58}_{-1.58}$ &   & 19 & h,e \\
\hline
0 & 0.2 & 0.1 & 5.69 & $^{+0.98}_{-0.85}$ & $^{+0.58}_{-0.47}$ & 52 & p \\
\hline
0.0 & 0.3 & 0.13 & 2.0 & $^{+0.7}_{-0.5}$ & $^{+0.5}_{-0.5}$ & 14 & e \\
\hline
0.02 & 0.2 & 0.14 & 3.43 & $^{+2.7}_{-1.6}$ & $^{+1.1}_{-0.6}$ & 4 & i,e \\
\hline
0.0 & 0.5 & 0.25 & 0 & $^{+2.4}_{-0.0}$ &   & 0 & b \\
\hline
0.2 & 0.3 & 0.25 & 1.7 & $^{+1.7}_{-1.7}$ &   & 1 & n \\
\hline
0.18 & 0.37 & 0.25 & 2.86 &   &   & 1 & j,e \\
\hline
0.3 & 0.4 & 0.35 & 5.3 & $^{+2.4}_{-2.4}$ &   & 5 & n \\
\hline
0.25 & 0.50 & 0.38 & 6.52 &   &   & 3 & k,e \\
\hline
0.2 & 0.6 & 0.4 & 6.9 & $^{+3.4}_{-2.7}$ & $^{+15.4}_{-2.5}$ & 3 & o \\
\hline
0.2 & 0.6 & 0.4 & 5.3 & $^{+3.9}_{-1.7}$ &   & 5.44 & a \\
\hline
0.4 & 0.5 & 0.45 & 7.3 & $^{+2.4}_{-2.4}$ &   & 9 & n \\
\hline
0.25 & 0.6 & 0.46 & 4.8 & $^{+1.7}_{-1.7}$ &   & 8 & l,e \\
\hline
0.2 & 0.6 & 0.47 & 8.0 & $^{+3.7}_{-2.7}$ & $^{+16.6}_{-2.6}$ & 8.8 & c \\
\hline
0.2 & 0.6 & 0.47 & 4.2 & $^{+0.6}_{-0.6}$ & $^{+1.3}_{-0.9}$ & 73 & d \\
\hline
0.25 & 0.85 & 0.55 & 5.4 & $^{+0.74}_{-0.66}$ & $^{+0.84}_{-0.82}$ & 37 & m \\
\hline
0.5 & 0.6 & 0.55 & 20.4 & $^{+3.8}_{-3.8}$ &   & 29 & n \\
\hline
0.6 & 0.7 & 0.65 & 14.9 &$^{+3.1}_{-3.1}$ &   & 23 & n \\
\hline
0.5 & 1.0 & 0.75 & 4.3 & $^{+3.6}_{-3.2}$ &   & 5.5 & b \\
\hline
0.7 & 0.8 & 0.75 & 17.8 &$^{+3.4}_{-3.4}$ &   & 28 & n \\
\hline
0.6 & 1.0 & 0.8 & 15.7 & $^{+4.4}_{-2.5}$ & $^{+7.5}_{-5.3}$ & 14 & o \\
\hline
0.6 & 1.0 & 0.8 & 9.3 & $^{+2.5}_{-2.5}$ &   & 18.33 & a \\
\hline
0.6 & 1.0 & 0.83 & 13.0 & $^{+3.3}_{-2.7}$ & $^{+7.3}_{-5.1}$ & 23.5 & c \\
\hline
0.87 & 1.27 & 0.9 & 9.32 &   &   & 5 & j,e \\
\hline
1.0 & 1.4 & 1.2 & 7.5 & $^{+3.5}_{-3.0}$ &   & 8.87 & a \\
\hline
1.0 & 1.4 & 1.2 & 11.5 & $^{+4.7}_{-2.6}$ & $^{+3.2}_{-4.4}$ & 6 & o \\
\hline
1.0 & 1.4 & 1.21 & 13.2 & $^{+3.6}_{-2.9}$ & $^{+3.8}_{-3.2}$ & 20.2 & c \\
\hline
1.0 & 1.5 & 1.25 & 10.5 & $^{+4.5}_{-5.6}$ &   & 10.0 & b \\
\hline
1.4 & 1.7 & 1.55 & 1.2 & $^{+5.8}_{-1.2}$ &   & 0.35 & a \\
\hline
1.4 & 1.8 & 1.6 & 4.4 & $^{+3.2}_{-2.5}$ & $^{+1.4}_{-1.1}$ & 2 & o \\
\hline
1.4 & 1.8 & 1.61 & 4.2 & $^{+3.9}_{-2.3}$ & $^{+1.9}_{-1.4}$ & 3.1 & c \\
\hline
1.5 & 2.0 & 1.75 & 8.1 & $^{+7.9}_{-6.0}$ &   & 3.0 & b \\
\end{longtable}
\footnotetext[3]{$10^{-5}$SNe $(Mpc/h_{70})^{-3}yr^{-1}$}
\footnotetext[4]{a$)$ \cite{kuznetsova08}, b$)$ \cite{poznanski07}, c$)$ \cite{dahlen08}, d$)$ \cite{neill06}, e$)$ \cite{blanc04}, f$)$ \cite{dilday08}, g$)$ \cite{cappellaro99}, h$)$ \cite{madgwick03}, i$)$ \cite{hardin00}, j$)$ \cite{galyam02}, k$)$ \cite{pain96}, l$)$\cite{tonry03}, m$)$ \cite{pain02}, n$)$ \cite{barris06}, o$)$ \cite{dahlen04}, p$)$ This paper}
\end{center}
\end{document}